\documentclass[sigconf]{acmart}
\usepackage[utf8]{inputenc} 
\renewcommand\footnotetextcopyrightpermission[1]{} 

\usepackage{graphicx}
\usepackage{caption}
\usepackage{subcaption}

\usepackage{color}
\usepackage{hyperref}
\usepackage{todonotes}

\usepackage{enumitem}

\usepackage{enumerate}
\usepackage{amsmath}
\usepackage{booktabs}
\usepackage{multirow}
\usepackage{mathrsfs} 

\acmDOI{}

\acmConference[FATREC'18]{2nd FATREC Workshop on Responsible Recommendation}{October 2018}{Vancouver, Canada}
\acmYear{2018}
\setcopyright{none}

\begin{document}

\title{Synthetic Attribute Data for Evaluating Consumer-side Fairness }

\author{Robin Burke}
\affiliation{
    \institution{DePaul University}
    \city{Chicago}
    \state{Illinois}
}\email{rburke@cs.depaul.edu}

\author{Jackson Kontny}
\affiliation{
    \institution{DePaul University}
    \city{Chicago}
    \state{Illinois}
}\email{jackson.kontny@gmail.com}

\author{Nasim Sonboli}
\affiliation{
    \institution{DePaul University}
    \city{Chicago}
    \state{Illinois}
}\email{nsonboli@depaul.edu}

\begin{abstract}

When evaluating recommender systems for their fairness, it may be necessary to make use of demographic attributes, which are personally sensitive and usually excluded from publicly-available data sets. In addition, these attributes are fixed and therefore it is not possible to experiment with different distributions using the same data. In this paper, we describe the Frequency-Linked Attribute Generation (FLAG) algorithm, and show its applicability for assigning synthetic demographic attributes to recommendation data sets.
  
\end{abstract}

\maketitle

\section{Introduction}

Fairness in recommender systems spans a number of different research questions. One key area is the impact of user demographic attributes on their experiences using recommender systems, an aspect of consumer-side fairness (C-fairness)~\cite{burke_robin_multisided_nodate}. In some application areas, such as employment, there may be a legal mandate to ensure that users in protected groups have similar quality recommendations to those who are not.

A challenge in performing C-fairness research is that demographic attributes are rarely included in public data sets used in recommendation, especially in sensitive areas such as employment: such attributes would make it much easier to de-anonymize the data and uncover the identities of the users.

In this position paper, we outline our solution, the Frequency-Linked Attribute Generation (FLAG) algorithm, for probabilistic generation of synthetic demographic attributes.

\subsection{Fairness}

A standard simplification in fairness-aware machine learning is to consider users as divided into protected and unprotected groups, where fairness towards the protected group is desired~\cite{zemel2013learning}. In the case of job seekers, the protected group may depend on the job category, but may often be associated with gender and / or racial / ethnic identity. 

For the 2017 RecSys Challenge~\cite{abel2017recsys}, the career-oriented social networking site XING\footnote{www.xing.com} released a data set consisting of interactions between users and job postings. Most attributes of jobs and users were anonymized, so that it is not possible to make use of any demographic information for fairness-aware recommendation research. The data set is large and sparse with over 10 million interactions. We produced a sample of the data by concentrating on users of career level 0 and within region 7, leaving approximately 410k users and around 3 million interactions. These users have profiles that range in length from one interaction up to 30 -- the very small number of users with larger profiles were removed.

\section{Synthetic attribute generation}

It is well established that different types of users have different behaviors in employment-seeking contexts~\cite{wanberg1996individuals}. Male job seekers tend to be more optimistic relative to expected salary, for example~\cite{heckert2002gender}. Similar differences are reported for race and ethnic identity, even when controlling for background~\cite{avery2003racial}. 

These findings suggest that the potential exists for a feedback loop in job recommendation with respect to job quality. White male users may click more optimistically on such jobs, and other users may be less likely to. A recommender system may pick up on this difference, and allocate recommendations accordingly, leading to an unacceptable degree of disparity between the quality of jobs presented to different groups. 
 
\subsection{Frequency-Linked Attribute Generation}

The FLAG algorithm does not attempt to uncover any ground truth about the demographic status of any individual or group within its input data, and could not be used for this purpose. We treat the task as one of generating a membership probability distribution, which can then be applied to assign a binary-valued attribute.

In order to serve as a useful proxy for unprotected / protected status, labeled $A$ and $B$, respectively, there are certain requirements that a synthetic demographic attribute should have:

\begin{itemize}
    \item Group labels should be assigned based on a probability distribution with every user having some non-zero probability of receiving either label $A$ or $B$.
    \item The feature should be correlated with differences in user behavior, so that it can be applied to data sets where only behavior is known.
    \item The data generator should be parameterized such that groups $A$ and $B$ can have different relative sizes, and that they can have behavioral profiles that vary in overall similarity. This will allow us to evaluate algorithms under a range of conditions.
\end{itemize}

To understand FLAG algorithm and how it meets these requirements, we start with the observation that profile sizes in recommendation data sets generally follow a power-law distribution with a small number of very active users and a much larger number of less active ones. Our subset of the XING data follows such a power law distribution for profile size with an estimated exponent of 1.45\footnote{Calculations performed with the powerRlaw package version 0.70.1 in R 3.4.4.}. Note that we are using profile length (number of clicks) as our behavioral indicator in this research, but it could be any other property that has a left-skewed distribution and that is probabilistically associated with a demographic attribute of interest.

Let $S$ be the distribution of profile sizes for users. $S(i)$ equals the number of users with profiles of size $i$. The maximum profile size is $k$: $k$ equals 30 in our XING subset. 
In FLAG, membership probability in groups $A$ and $B$ is a function of the size of a user's profile, following a power law distribution. 

We begin by setting the probability of membership in group B to be $f_B(i) = 1/i^\alpha$, with the $\alpha$ parameter controlling the skew of the distribution. As $\alpha$ approaches $0$, the distribution approaches a uniform line at 1 -- all users are in group $B$ -- and as $\alpha$ approaches $\infty$, it approaches zero -- all users are in group $A$. 

This gives us a variety of different distributional shapes for groups $A$ and $B$, but it does not allow us to control their relative sizes. The expected number of group $B$ users under $f$ is given by:

\begin{equation}
E_f(|B|) = \sum_{i=1}^{k} S(i) f_B(i)
\end{equation}

To control the relative sizes of the two groups, we introduce a parameter $\beta$ that specifies the fraction of users that we would  in group $B$. In other words, $E(|B|) = \beta |U|$. We can achieve this result by uniformly scaling each $f_B(i)$ value. That is, we multiply each $f_B(i)$ by $\beta |U|/E_f(|B|)$. This ensures that each group has, in expectation, the desired size. Taking both parameters into account, we can write the FLAG function as a membership probability distribution over profile sizes: 
\begin{equation}
    \text{FLAG}_B(j) = \frac{\beta |U|}{j^\alpha \sum_{i=1}^{k} S(i) /i^\alpha}
\end{equation}.

To generate attributes, we process each user profile $u$ and given the profile size $i$, we calculate the group $B$ membership probability $p=\text{FLAG}_B(i)$. We conduct a Bernoulli trial with probability $p$ and on success, assign $u$ to group $B$; otherwise, group $A$. 

Note that not all combinations of $\alpha$ and $\beta$ are possible. If we attempt to make the two groups very different in behavior (with a large $\alpha$ value), it may be impossible to have the groups be similar in size. Legal $\beta$ values will be in the following range:

\begin{equation}
    0 < \beta \leq \frac{E_f(|B|)}{|U|*f_B(1)} = \frac{E_f(|B|)}{|U|}
\end{equation}

\subsection{Generation Results}

\begin{figure}
\includegraphics[width=\columnwidth]{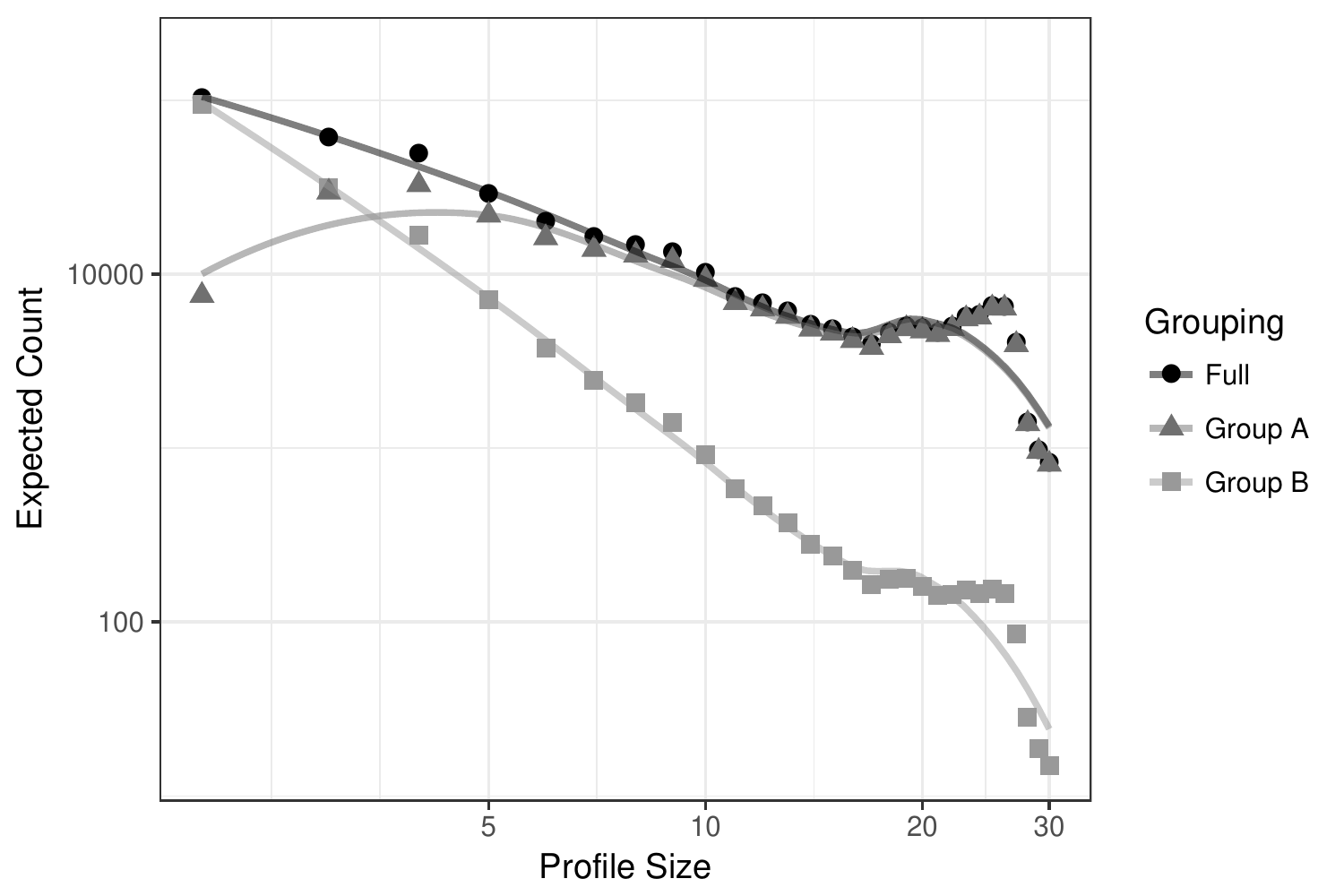}
\caption{Distributions for generated groups.}
\label{fig:ab-loglog}
\end{figure}

In Figure~\ref{fig:ab-loglog}, we have set $\alpha=1.45$ to mirror the overall distribution and $\beta$ to 0.4. (Note that with $\alpha=1.45$, the maximum $\beta$ value is 0.43 for the XING data set.) The figure shows the distribution of the expected value of the profile count at each size, using a log-log scale. The full profile size distribution is included for comparison. As can be seen, group $A$ dominates in the lower profile sizes where the bulk of the data lies. As the proportion of group $B$ nodes gets smaller and smaller, the group $A$ distribution approaches the original data.

\begin{figure}
\includegraphics[width=\columnwidth]{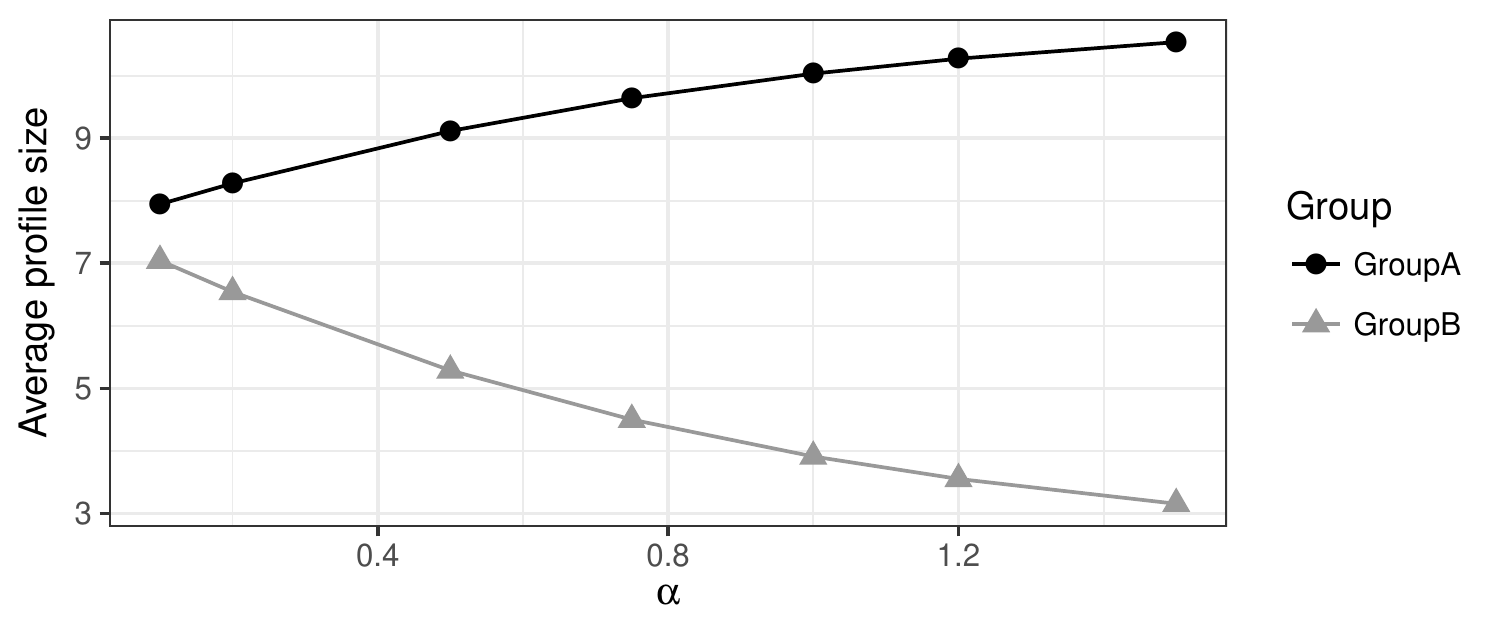}
\caption{Expected profile sizes with increasing $\alpha$.}
\label{fig:alpha-plot}
\end{figure}

Figure~\ref{fig:alpha-plot} shows legal values for $\alpha$ with $\beta=0.4$. As $\alpha$ increases, the behaviors of the two groups, as expressed in profile size, become increasingly different. 

\section{Examples}

\begin{figure}
\includegraphics[width=\columnwidth]{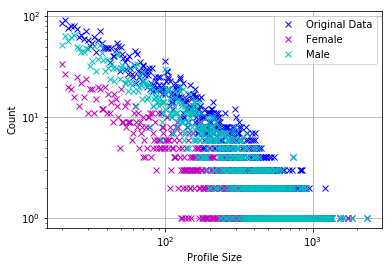}
\caption{Profile length distribution by gender.}
\label{fig:ml-gender}
\end{figure}

Recommendation data sets for public use that contain sensitive demographic characteristics are relatively rare. To demonstrate the validity of our synthetic data generation method, we worked with the well-known MovieLens 1M data set, which does contain a gender attribute for users. Figure~\ref{fig:ml-gender} shows the distribution of profile lengths for this data. We can see that female users make up a minority of the user base (1709 females vs. 4331 males) and they tend to have shorter profile lengths (average of 164 movies per male users, shown in light blue, and 144 movies per female user, in magenta). The total distribution is dark blue. Note the relatively linear appearance of the distribution in the log-log plot, suggesting that a power law is an appropriate model.

\begin{figure}
\includegraphics[width=\columnwidth]{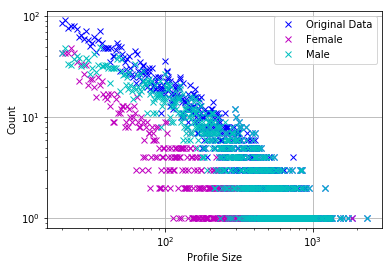}
\caption{Profile length distribution by synthetic attribute. }
\label{fig:ml-gender-synth}
\end{figure}

We followed the procedure described above to generate a synthetic attribute associated with profile length. We tuned the $\alpha$ and $\beta$ parameters to match the real distribution as closely as we could, arriving at $\alpha=0.23$ and $\beta=0.34$. Figure~\ref{fig:ml-gender-synth} shows one run of attribute generation using these values, and indicates that the distribution of the $A/B$ feature matches in many respects the gender feature in the original data. The total number of group $A$ profiles is 4592 in this run and 1468 in group $B$. Since it is a stochastic process, different runs produce slightly different results.

\begin{figure}
\includegraphics[width=\columnwidth]{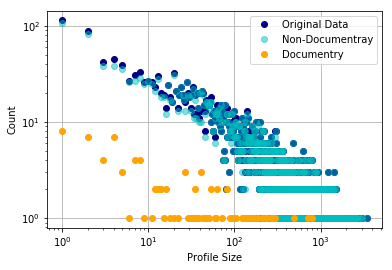}\caption{Profile length distribution by ``Documentary'' genre. }
\label{fig:movie_loglog}
\end{figure}

We can apply a similar process to features associated with items, which would be needed for the evaluation of provider-side fairness (P-fairness) in situations where the demographics of providers were relevant. For example, jobs in minority-owned businesses might be considered a protected class in some job recommendation settings. Again, we turn to MovieLens using genre information as the protected feature. Figure~\ref{fig:movie_loglog} shows the distribution of the item profiles for movies with the ``Documentary'' feature as opposed to those without this feature. There are considerably fewer such movies (110 out of 3706) and they tend to have much smaller item profiles -- meaning that documentaries does not tend to attract as many ratings as other movies in the data set. 

\begin{figure}
\includegraphics[width=\columnwidth]{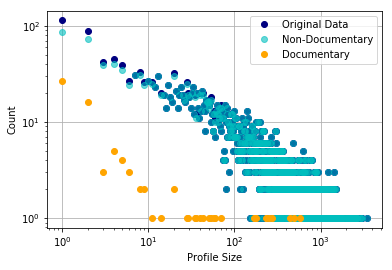}
\caption{Profile length distribution by synthetic genre attribute. }
\label{fig:synth_loglog_movie}
\end{figure}

For the data set, we generated a synthetic attribute with $\alpha=0.3$ and $\beta=0.10$, as shown in Figure~\ref{fig:synth_loglog_movie}. In this case, the fit is not quite as good, somewhat fewer group $B$ movies, not extending quite as much in profile length as the original data. Compared to the original attribute distribution, we see that the slope of the distribution is too steep. However, further adjustment of $\alpha$ produces illegal $\beta$ values. Such findings suggest that it may be necessary to augment the model with a third parameter, adjusting the power-law baseline value to account for all distributions that may arise.

\section{Ethical Considerations}

As noted earlier, the beneficial aim of this research is to enable experimental development of recommendation algorithms with improved fairness properties, an important goal given the prevalence of recommendation algorithms in commerce, social media, and other online settings. Without synthetic data, research into such systems becomes limited by the availability of data about real individuals and their sensitive demographic characteristics. A select group of researchers (especially in industry settings) may have access to such data, but the inability to share data and compare results inevitably slows research productivity and innovation. Synthetic data is therefore essential to progress towards fairness-aware recommender systems.

Two concerns might be raised about the ethical propriety of generating synthetic data for recommendation evaluation: de-anon\-ym\-ization and external validity. De-anonymization occurs when operations performed on the data make it possible to recover some aspects of it that were not intended to be made public: for example, the gender of the user, if this information was withheld in the original data release, or most significantly, the identity of a system user. Since recommendation data sets contain profiles of user activity including consumer behavior, the relevation of user identity is considered a important risk. Data sets such as those associated with the RecSys Challenge are carefully anonymized precisely so such recovery of individual identity is not possible.

Note first that the FLAG algorithm does not attribute demographic attributes to users. It generates a synthetic \textit{A} or \textit{B} label, which does not have a demographic meaning. We do know, based on the way in which the labels are generated, that a user assigned to group \textit{A} is more likely to have a larger user profile, but that is only a probabilistic association. We assume that profile length is distributed according to a power law, and in such a distribution, there will always be more users with small profiles even in group \textit{A}. It should also be noted that every users' profile length is readily derived in any recommendation data set. Therefore, access to the output of FLAG tells a researcher nothing about an anonymized user record that could not already be determined by simple inspection of the input data to see the position of that user's profile length relative to the distribution as a whole.

The question of external validity asks whether demographic attributes of interest such as gender, race, age, etc. follow the type of distribution we assume and are linked to profile length as our model suggests. If such attributes have a very different relation to the input variable, then our synthetic attribute will fail to be a good stand-in for the real demographic attribute. The example of MovieLens above suggests that the labels generated are statistically similar to some user and item features, but this association would have to be verified in any domain where this technique is to be applied.

Since FLAG assigns labels probabilistically, it will not capture other aspects of the data that may be correlated with a demographic attribute. For example, our prior work demonstrated some differences in genre preferences between male and female users in the MovieLens 1M data set~\cite{burke2018balanced}. The group \textit{A} and group \textit{B} users assigned by FLAG would not show these differences. This is a consequence of using a single dimension of user behavior to control attribute generation. Note that other features such as user age might also be associated with differences in profile length and any such variables would be conflated in synthetic attribute production. This is one reason to avoid any claim that FLAG is inferring unknown demographic aspects of users. 

In order to make the assigned labels track additional aspects of user behavior, such as genre preference, these dimensions would have to be incorporated into the model. The model would then move closer towards an inferential approach (inferring missing demographic attributes) rather than a synthetic one in which the labels are meaningless although systematically assigned. This tension can be resolved by generating fully synthetic data, with an algorithm that generates all the profile information and captures demographic differences as well. This is much more complex challenge that we leave for future work.

We believe that using the fairly neutral and domain-general profile length characteristic is a good compromise between the concerns of avoiding attribute inference and of generating unrealistic data. However, the question of the external validity of FLAG's attribute generation remains to be fully answered. It is possible that fairness results relative to synthetic data will not translate to real-world applications. This concern must be answered through additional research. 

\section{Conclusion}

Fairness-aware recommendation research requires appropriate data for experimentation. However, sensitive demographic characteristics that are of most interest in areas where fairness is important are precisely those that are least likely to be disclosed. This paper has outlined the Frequency-Linked Attribute Generation (FLAG) algorithm for generating such attributes. We show it is possible to augment real user data with synthetic data designed to closely match the characteristics of the real attributes in distribution and linkage to user behavior. We provide a suggestive example of generating synthetic data in the MovieLens data set and show that we are able to reproduce a probabilistic association between a demographic attribute and profile length.

\bibliographystyle{ACM-Reference-Format}
\bibliography{main}

\end{document}